\documentclass[]{JHEP}
\usepackage{epsfig,mathbbol,verbatim}

\def\D{\mbox{D}}
\def\d{\mbox{d}}

\title{Hopf defects as seeds for monopole loops}

\author{Falk Bruckmann\\
Friedrich Schiller University Jena\\
Theoretisch-Physikalisches Institut\\
Max-Wien-Platz 1, D-07743 Jena\\
\email{brk@tpi.uni-jena.de}
}

\abstract{We investigate the relation between instantons and monopoles
in the Laplacian Abelian Gauge using analytical methods in the
continuum.
Our starting point is the fact that the 't~Hooft instanton with its
high symmetry leads to a pointlike defect with Hopf invariant one.
In order to generalise this result we partly break the symmetry by a
local perturbation.
We find that for generic configurations near the
't~Hooft instanton the defects become loops.
The analytical results show explicitly that these defects are magnetic
monopoles with unit charge.
In addition, the monopoles are twisted to account for the
instanton number of the background.}


\preprint{FSU TPI 09/00\\hep-th/0011249.}

\begin{document}

\section{Introduction}

Topological objects are prominent examples of non-perturbative effects
in quantum field theories. For Yang-Mills theories 
these are instantons, magnetic monopoles (and center vortices),
respectively. While the first are intimately connected to the gauge
invariant topological density
and responsible for
chiral symmetry breaking \cite{schaefer:98},
the others are visible only after an Abelian 
(center) gauge fixing \cite{thooft:81a,deldebbio:97,alexandrou:00a},
and supposed to be responsible for confinement.
Since both physical effects take
place below the same critical temperature \cite{kogut:83}, a relation between
instantons and monopoles is highly desirable but still not fully known.
The first result in this direction is due to Rossi \cite{rossi:79}:
a static 't~Hooft-Polyakov monopole can be built out
of an array of instantons placed along the time axis.
A similar construction exists for the caloron \cite{kraan:98a}.

To see how instantons are built from monopoles we take the point of view
of Abelian projections (for a detailed prescription see
\cite{bruckmann:00c}).
An Abelian gauge is a partial gauge fixing
leaving the maximal Abelian subgroup\footnote{
We will restrict ourselves to the gauge group $SU(2)$,
where the maximal Abelian subgroup is simply
$U(1)$ embedded in terms of diagonal matrices.}
untouched.
The needed gauge
transformation is best described by the diagonalisation of an `auxiliary Higgs
field' $\phi$ in the adjoint representation. Defects occur where this
field vanishes, i.e.~the diagonalisation becomes ambiguous.
Since this means solving three equations, generic defects in four
dimensions form lines.
Moreover, the normalised Higgs field $n=\phi/|\phi|$
perpendicular to those lines generically is a hedgehog.
It can be diagonalised only at the expense of introducing
a singular gauge field, the Dirac monopole.
By charge conservation defects form closed lines, i.e.~loops.

Topological arguments enforce the existence of defects for
every configuration with non-vanishing instanton number on the
four-sphere\footnote{Such a strong statement does not hold for the
four-torus as is plausible from the existence of {\em Abelian instantons}
\cite{thooft:81c}.}. 
The topological properties of the defects necessary to generate an
instanton number are highly non-trivial for general Abelian gauges \cite{jahn:00}.
However, the relation between instantons and {\em static}
monopoles in the Polyakov gauge is well understood
\cite{weiss:81,griesshammer:97,reinhardt:97b,ford:98,jahn:98,ford:99a,ford:99b}.

Much less is known analytically about defects in the Laplacian Abelian
gauge (LAG) \cite{vandersijs:97,vandersijs:98,vandersijs:99}.
The Higgs field of the LAG 
is defined as the lowest eigenvector of the gauge covariant Laplacian
in the background of $A$,
$-\D^2[A]\phi=E_0\phi$.
The LAG has turned out to be `very useful' on the lattice in the sense
that it shares Abelian dominance with the
Maximal Abelian Gauge (MAG) \cite{ilgenfritz:00}
but does not suffer from a severe
Gribov problem \cite{bali:96,deforcrand:01b}.
Monopole loops have been observed for instanton backgrounds in the LAG
by numerical means \cite{reinhardt:00,deforcrand:01a}.
A fully analytical treatment, however, is very difficult.
So far it was only possible for the
't~Hooft instanton \cite{bruckmann:01a} (and the meron \cite{reinhardt:00})
which is highly symmetric and thus non-generic; the defect is
degenerate to a point and localised at the instanton core (see below).

The present work is the first step towards an analytical investigation
of generic configurations in the LAG.
By breaking the high symmetry, we show that for configurations near the 't~Hooft
instanton the defect manifold becomes a loop (even a circle, Section \ref{first_section}).
 Furthermore, the associated hedgehog is twisted once along the
loop (the simplest possibility to account for the instanton number, Section \ref{second_section}).
As a by-product, the picture of Hopf defects as `monopole loops
with vanishing radius' is proven.

\section{From Hopf defects to monopole loops}
\label{first_section}

The ground state of the $SU(2)$ covariant Laplacian in the background
of a 't~Hooft instanton in regular gauge 
is of the form
\cite{bruckmann:01a},
\begin{eqnarray}
\label{phi_instanton}
\phi=f(r)n_{\rm H},\qquad f(r)\stackrel{r\rightarrow
0}{\longrightarrow}r^2,
\end{eqnarray}
where $n_{\rm H}$ is the standard Hopf map\footnote{being the projection 
in the Hopf bundle
describing the Dirac monopole \cite{ryder:80}} \cite{hopf:31,nakahara:90,dubrovin:85}
\begin{eqnarray}
\label{hopf_map}
n_{\rm H}\equiv\left(\begin{array}{c}
2(\hat{x}_1\hat{x}_3+\hat{x}_2\hat{x}_4) \\
2(\hat{x}_2\hat{x}_3-\hat{x}_1\hat{x}_4) \\
\hat{x}_1^2+\hat{x}_2^2-\hat{x}_3^2-\hat{x}_4^2
\end{array}\right),\qquad\hat{x}_\mu\equiv x_\mu/r
\end{eqnarray}
This Higgs field $\phi$ vanishes quadratically\footnote{in agreement with lattice simulations
\cite{deforcrand:xx}} at the
instanton core -- the origin -- where a pointlike defect is located.
We conjecture that this behaviour is not a feature
of the particular Abelian gauge chosen, but rather a matter of symmetry. The
't~Hooft instanton is spherically symmetric (in a proper definition
involving gauge transformations); hence any Abelian gauge which
does not break the rotational symmetry $SO(4)$ enforces the Higgs
field to be spherically symmetric as well. Monopole loops would break 
this symmetry, while pointlike defects (as well as $S^3$ defect
manifolds) do not.

On the other hand it is very easy to verify that the Higgs field
(\ref{hopf_map}) has the right topological behaviour.
Living in an associated bundle it
must have the same boundary conditions\footnote{in the bundle language
the same transition functions} as the gauge field,
\begin{eqnarray}
r\rightarrow\infty:\qquad A\rightarrow ig\d g^\dagger,\quad
n\rightarrow g\,\mbox{const}\,g^\dagger
\end{eqnarray}
For the instanton in regular gauge we have $g=h\equiv\hat{x}_4\Eins_2+i\hat{x}_a\sigma_a$.
It follows that $n$, being a mapping from $S^3$ (in coordinate space) to
$S^2$ (in color space), must have a Hopf invariant equal to the
instanton number (equal to the winding of $h$).
$n_{\rm H}=h\,\sigma_3/2\,h^\dagger$ is just
the prototype mapping with Hopf invariant one.

Let us now {\em slightly perturb} the 't~Hooft instanton, $A=A_{\rm
inst}+\lambda\delta A$, with perturbation parameter $\lambda$.
The usual Schr\"odinger perturbation theory for
the change of the groundstate, $\phi=\phi_{\rm inst}+\lambda\delta\phi$,
requires access to all eigenvalues and
eigenfunctions of $-\D^2[A_{\rm inst}]$. These are not known
analytically. But if perturbation theory is valid, the size of the
defect manifold (i.e.~the size of the expected monopole loop) is small.
Therefore we can restrict ourselves to the vicinity of the origin.
There we can Taylor expand $\delta\phi$;
for our purposes even the lowest order approximation is sufficient.
Thus the Higgs field of a generic configuration $A$
\textit{close to the
instanton} (in orbit space) and \textit{near the origin} (in
coordinate space) is (cf. (\ref{phi_instanton})),
\begin{eqnarray}
\label{phi_perturbed_1}
\phi=\phi_{\rm inst}+\lambda\delta\phi=
r^2\,n_{\rm H}+R^2\:\mbox{const}
\end{eqnarray}
where we have introduced a radius parameter $R$, since the Higgs field
in our convention is of dimension (length)$^2$.

Without loss of generality
we specialise to a perturbation pointing in the third color
direction,
\begin{eqnarray}
\label{phi_perturbed_2}
\phi=r^2\,n_{\rm H}-R^2\left(\begin{array}{c}
0\\0\\1
\end{array}\right)
\end{eqnarray}
all other cases can be obtained by rotations.
A straightforward calculation shows that the zeros of $\phi$ are then
on the {\em circle} $C:\:\:x_1^2+x_2^2=R^2,\:x_3=x_4=0$.  
Its size scales with the perturbation parameter $R=\sqrt{\lambda}$
(see (\ref{phi_perturbed_1})).
The perturbation has {\em enlarged the defect manifold from a point to
a loop}, thereby breaking the spherical symmetry.
Such a picture was conjectured in
\cite{brower:97b} for instantons in the MAG, but there
the formation of the loop is suppressed by the gauge fixing functional
\cite{bruckmann:00a}.

Notice that the deformed Higgs field $\phi$ is now on a different
orbit, since we changed its zeros which are gauge invariant.
Its global properties, however, remain the
same (as we will also see in the next section) because we performed only
a local perturbation around the origin.

\section{Monopole charge and twist}
\label{second_section}

To proceed further we introduce polar angles for both coordinate space
$\mathbb{R}^4$, 
\begin{eqnarray}
x=(r_{12}\cos\varphi_{12},
r_{12}\sin\varphi_{12},
r_{34}\cos\varphi_{34},
r_{34}\sin\varphi_{34}),
\: r_{12}=r\cos\vartheta,\,r_{34}=r\sin\vartheta
\end{eqnarray}
and color space $\mathbb{R}^3$ resp.~$S^2$,
\begin{eqnarray}
n=\left(\begin{array}{c}
\sin\beta\cos\alpha \\
\sin\beta\sin\alpha \\
\cos\beta
\end{array}\right).
\end{eqnarray}
The Hopf map (\ref{hopf_map}) is given by assigning
\begin{eqnarray}
\label{angles_hopf}
\alpha_{\rm H}=\varphi_{12}-\varphi_{34},\qquad
\beta_{\rm H}=2\vartheta=\arctan\frac{2r_{12}r_{34}}{r_{12}^2-r_{34}^2}
\end{eqnarray}
while the perturbation (\ref{phi_perturbed_2}) corresponds to a
deformation of $\beta$,
\begin{eqnarray}
\label{angles_monopole}
\alpha=\varphi_{12}-\varphi_{34},\qquad
\beta=\arctan\frac{r^2\sin(2\vartheta)}{r^2\cos(2\vartheta)-R^2}
=\arctan\frac{2r_{12}r_{34}}
{r_{12}^2-r_{34}^2-R^2}
\end{eqnarray}
It turns out that this
Higgs field perfectly agrees with the one considered in
\cite{brower:97b,jahn:00},
$\beta=\vartheta_++\vartheta_-,\:
\tan\vartheta_{\pm}=r_{34}/(r_{12}\pm R)$.

\FIGURE[b]{
\begin{minipage}{0.45\linewidth}
\begin{center}
\epsfig{figure=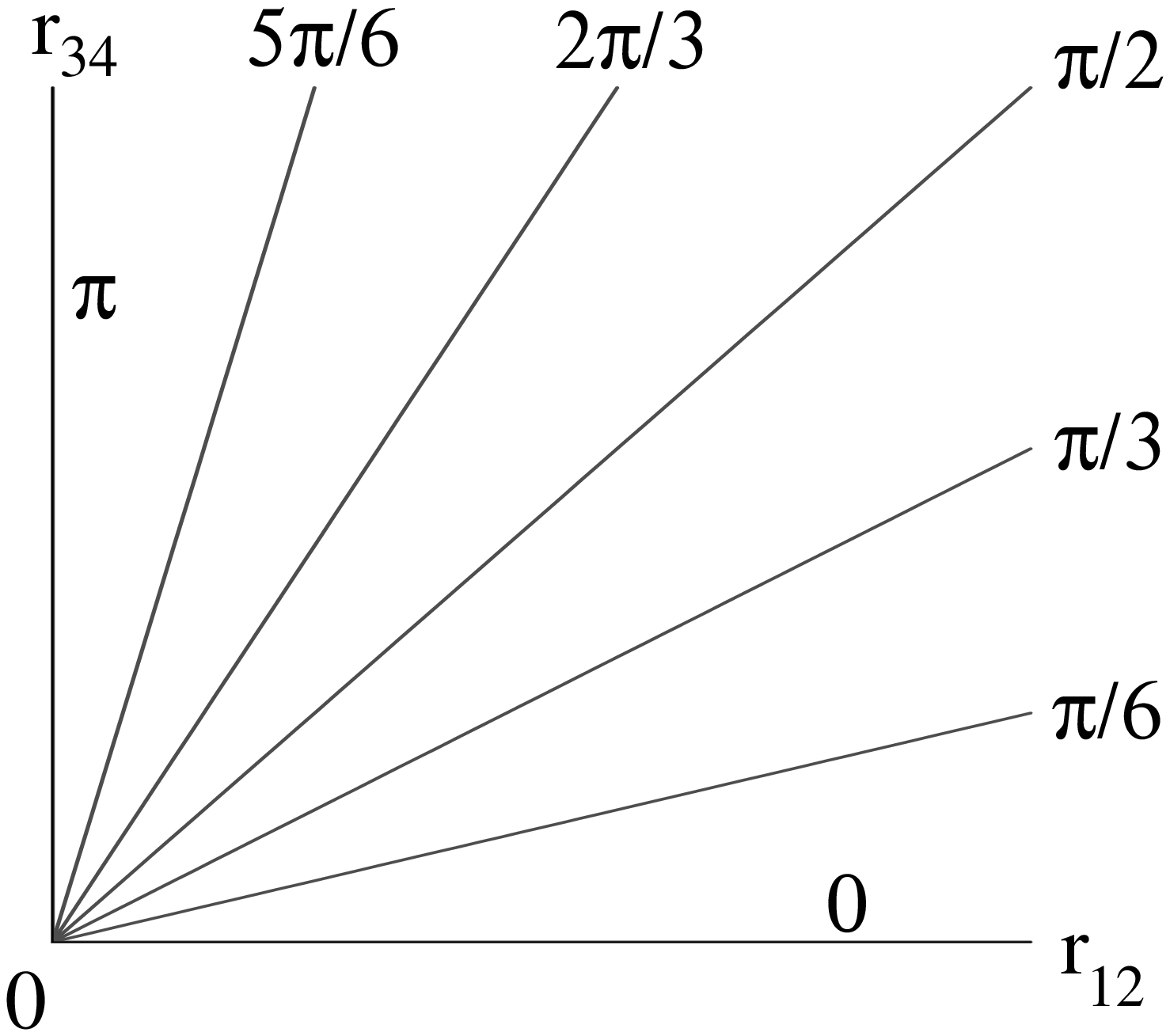,scale=0.35}
\end{center}
\end{minipage}
\begin{minipage}{0.45\linewidth}
\begin{center}
\epsfig{figure=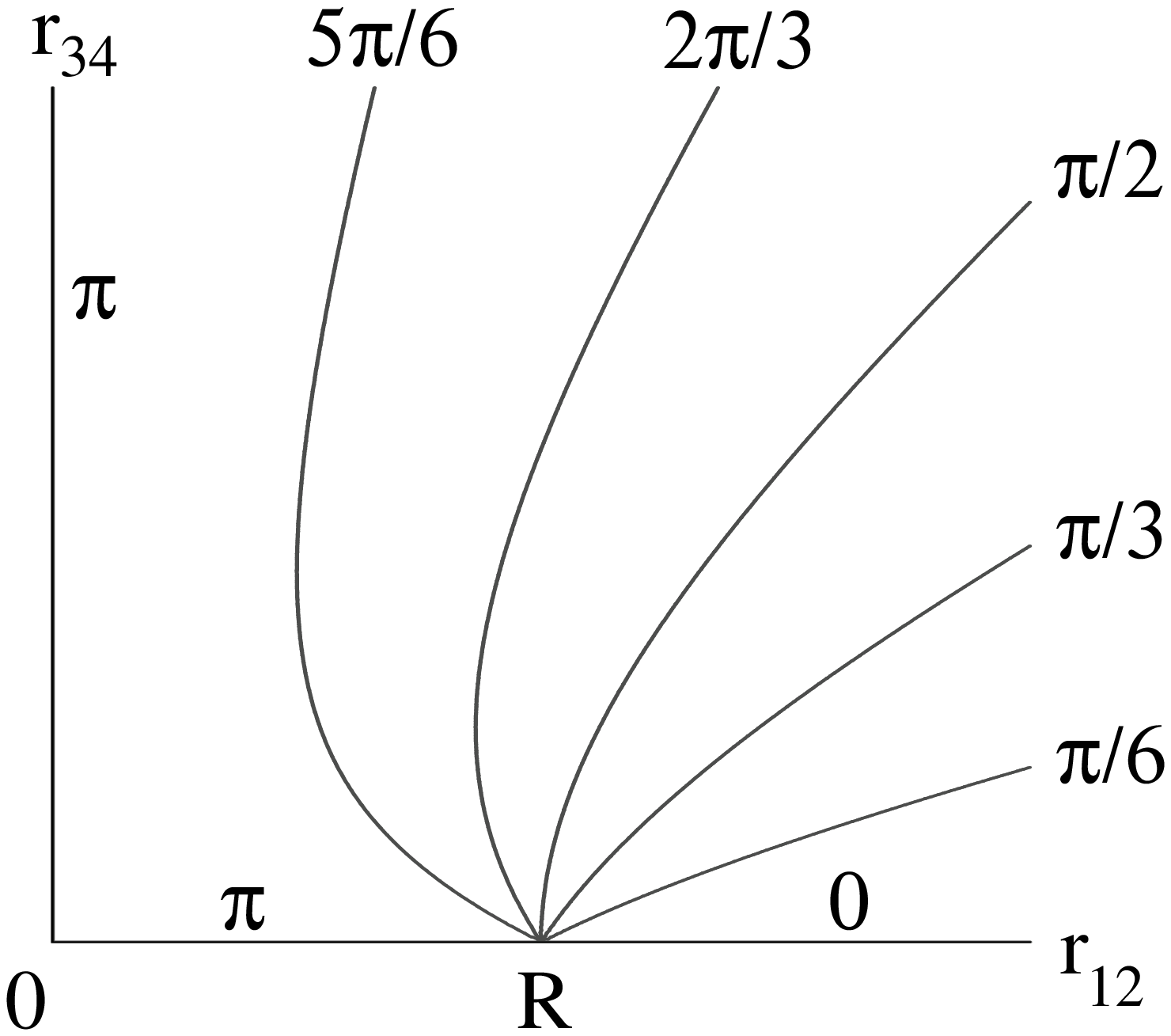,scale=0.35}
\end{center}
\end{minipage}
\caption{Lines of constant $\beta$ as a function of the two
radial coordinates $r_{12}$
and $r_{34}$ for the Hopf defect (left, cf.~(\ref{angles_hopf})) and 
a monopole loop at $r_{12}=R,\:r_{34}=0$ (right, essentially copied 
from \cite{brower:97b,jahn:00}, cf.~(\ref{angles_monopole})).  
In both cases the remaining polar angle is $\alpha=\varphi_{12}-\varphi_{34}$.}
\label{process}
}

From Fig.~\ref{process} it is obvious that both Higgs fields
(\ref{angles_hopf}) and (\ref{angles_monopole}) agree
in their global ($r\rightarrow\infty$) properties.
For the local properties of the new field it is important to notice
that on the loop $C$ both angles $\alpha$ and
$\beta$ are singular.
In a vicinity perpendicular to the loop they
take on all values $\alpha\in[0,2\pi],\:\beta\in[0,\pi]$.
Put differently, the normalised field $n$ is homotopic to the hegdehog: on any
two-sphere perpendicular to the
loop it covers the whole two-sphere in color space exactly once.
Viewed as a mapping $n:\:S^2\rightarrow S^2$ it has winding number one.
From the 't~Hooft-Polyakov monopole it is well-known that such a Higgs
field cannot be diagonalised\footnote{or brought to `unitary gauge'}
smoothly \cite{arafune:75}.  

One may nevertheless diagonalise $n$ by a singular gauge
transformation; in this way the Dirac monopole appears in the gauge
field. The relevant gauge transformation is
\begin{eqnarray}
\label{gauge_transformation}
g=e^{i\gamma\sigma_3/2}e^{i\beta\sigma_2/2}e^{i\alpha\sigma_3/2}
\end{eqnarray}
The residual $U(1)$-freedom (of rotations around the third color axis)
is encoded in $\gamma$. We choose
\begin{eqnarray}
\gamma=\varphi_{12}+\varphi_{34}
\end{eqnarray}
for which $g$ is singular on the disc $D:\:\:r_{12}\leq R,\:r_{34}=0$,
spanned by the loop $C$.

The gauge transformation (\ref{gauge_transformation}) induces an
inhomogeneous term, the Abelian part of which is
\begin{eqnarray}
a\equiv(i\Omega\d\Omega^{-1})_3=
\d\gamma+\cos\beta\d\alpha
\end{eqnarray}
One can easily compute the Abelian field strength,
\begin{eqnarray}
f&\equiv&\d a=f^{\rm reg}+f^{\rm sing},\\
f^{\rm reg}&=&-\sin\beta\,\d\beta\wedge\d\alpha,\\
f^{\rm sing}&=&(1-\cos\beta)\,\d^2\varphi_{34}
=4\pi\theta(R-r_{12})\delta(r_{34})\d x_3\wedge\d x_4
\end{eqnarray}
The regular part $f^{\rm reg}$ is just the Coulombic magnetic field,
while the singular part $f^{\rm sing}$ is the set of all Dirac strings
filling the disc $D$, called the Dirac sheet \cite{dirac:48}.
We use the latter to identify the monopoles as endpoints of Dirac
strings. The magnetic current is,
\begin{eqnarray}
k\equiv *\d f^{\rm sing}=4\pi\delta(r_{12}-R)\delta(r_{34})\d\varphi_{12}
\end{eqnarray}

The angle $\alpha$ not only depends on $\varphi_{34}$ (which gives the
hedgehog) but also on the
world-line coordinate $\varphi_{12}$. This means that the monopole is `twisted'
\cite{taubes:84a} once: the Higgs field $n$
rotates once around the third axis in isospace while moving 
along the loop. Our Higgs field is such that the complicated relation
in \cite{jahn:00} reduces to
\begin{eqnarray}
\mbox{Hopf invariant}=\mbox{magnetic charge}\times\mbox{twist},\qquad 1=1\times 1.
\end{eqnarray}

Gradually `switching off' the perturbation $\lambda\delta\phi$
one can see that {\em a Hopf defect emerges when
a twisted monopole loop is shrunk to vanishing radius}.
The Dirac sheet $D$ degenerates to a point, too.
Notice that the two-spheres used to measure the
magnetic charge are incapable to detect the Hopf defect.
Instead one has to pass to three-spheres surrounding a point in four dimensions.

\section{Conclusions}

We have investigated the Laplacian Abelian Gauge in the vicinity of
the 't~Hooft instanton by means of Schr\"odinger perturbation theory.
While the spherically symmetric instanton is related to a pointlike
defect, a generic (constant) perturbation induces a monopole loop with unit
charge and twist (cf.~(\ref{angles_monopole}) and its interpretation). 
Together these topological quantities give rise to the Hopf invariant,
which reflects the instanton number of the background gauge field.
Extending the correlation between the defect manifold and the
instanton core in the unperturbed case,
the instanton density of the new background is supposed to
be localised on a circle \cite{garciaperez:00,hansen:xx}.
Our result implies that for isolated, highly symmetric instantons
the monopole loops are very small.
Such small loops could be missed in lattice simulations of
the Abelian and monopole string tension.

Since isolated instantons are not sufficient for confinement, the
defects induced by them cannot be the whole story.
Bringing more instantons and anti-instantons close 
to each other, monopole loops start to spread out
\cite{brower:97b,reinhardt:00}.
Percolation is achieved if there are 
monopole loops extending over the whole space-time
as verified in lattice simulations \cite{bornyakov:92,ivanenko:93}.
Further analytic approaches are necessary to better understand this mechanism.

\acknowledgments{
The author thanks D\"orte Hansen, Thomas Heinzl and Andreas Wipf
for helpful discussions and a careful reading of the
manuscript. Furthermore he is grateful
to Manuel Asorey, Philippe de Forcrand and Michele Pepe for
their hospitality and sharing their insights.
}

\bibliographystyle{../../JHEP}
\bibliography{}

\providecommand{\href}[2]{#2}\begingroup\raggedright\begin{thebibliography}{10}

\bibitem{schaefer:98}
T.~Sch{\"a}fer and E.~V. Shuryak, {\it Instantons in QCD}, {\em
  Rev.~Mod.~Phys.} {\bf 70} (1998) 323,
  [\href{http://xxx.lanl.gov/abs/hep-ph/9610451}{{\tt hep-ph/9610451}}].

\bibitem{thooft:81a}
G.~'t~Hooft, {\it Topology of the gauge condition and new confinement phases in
  non-Abelian gauge theories},  {\em Nucl.~Phys.} {\bf B190} (1981) 455.

\bibitem{deldebbio:97}
L.~{Del~Debbio}, M.~Faber, J.~Greensite, and \v{S}. Olejnik, {\it Center
  dominance and Z(2) vortices in SU(2) lattice gauge theory},  {\em Phys.~Rev.}
  {\bf D55} (1997) 2298, [\href{http://xxx.lanl.gov/abs/hep-lat/9610005}{{\tt
  hep-lat/9610005}}].

\bibitem{alexandrou:00a}
C.~Alexandrou, M.~D'Elia, and P.~de~Forcrand, {\it The relevance of center
  vortices},  {\em Nucl.~Phys.~Proc.~Suppl.} {\bf 83} (2000) 437--439,
  [\href{http://xxx.lanl.gov/abs/hep-lat/9907028}{{\tt hep-lat/9907028}}].

\bibitem{kogut:83}
J.~Kogut, M.~Stone, H.~W. Wyld, W.~R. Gibbs, J.~Shigemitsu, S.~H. Shenker, and
  D.~K. Sinclair, {\it Deconfinement and chiral symmetry restoration at finite
  temperatures in SU(2) and SU(3) gauge theories},  {\em Phys.~Rev.~Lett.} {\bf
  50} (1983) 393.

\bibitem{rossi:79}
P.~Rossi, {\it Propagation functions in the field of a monopole},  {\em
  Nucl.~Phys.} {\bf B149} (1979) 170.

\bibitem{kraan:98a}
T.~C. Kraan and P.~van Baal, {\it Periodic instantons with non-trivial
  holonomy},  {\em Nucl.~Phys.} {\bf B533} (1998) 627--659,
  [\href{http://xxx.lanl.gov/abs/hep-th/9805168}{{\tt hep-th/9805168}}].

\bibitem{bruckmann:00c}
F.~Bruckmann and G.~'t~Hooft, {\it Monopoles, instantons and confinement},
  \href{http://xxx.lanl.gov/abs/hep-th/0010225}{{\tt hep-th/0010225}}.

\bibitem{thooft:81c}
G.~'t~Hooft, {\it Some twisted selfdual solutions for the Yang-Mills equations
  on a hypertorus},  {\em Comm.~Math.~Phys.} {\bf 81} (1981) 455.

\bibitem{jahn:00}
O.~Jahn, {\it Instantons and monopoles in general abelian gauges},  {\em
  J.~Phys.} {\bf A33} (2000) 2997--3019,
  [\href{http://xxx.lanl.gov/abs/hep-th/9909004}{{\tt hep-th/9909004}}].

\bibitem{weiss:81}
N.~Weiss, {\it Effective potential for the order parameter of gauge theories at
  finite temperature},  {\em Phys.~Rev.} {\bf D24} (1981) 475.

\bibitem{griesshammer:97}
H.~W. Griesshammer, {\it Magnetic defects signal failure of Abelian projection
  gauges in QCD},  \href{http://xxx.lanl.gov/abs/hep-ph/9709462}{{\tt
  hep-ph/9709462}}.

\bibitem{reinhardt:97b}
H.~Reinhardt, {\it Resolution of Gauss' law in Yang-Mills theory by gauge
  invariant projection: Topology and magnetic monopoles},  {\em Nucl. Phys.}
  {\bf B503} (1997) 505, [\href{http://xxx.lanl.gov/abs/hep-th/9702049}{{\tt
  hep-th/9702049}}].

\bibitem{ford:98}
C.~Ford, U.~G. Mitreuter, J.~M. Pawlowski, T.~Tok, and A.~Wipf, {\it Monopoles,
  polyakov loops and gauge fixing on the torus},  {\em Ann.~Phys.~(N.Y.)} {\bf
  269} (1998) 26, [\href{http://xxx.lanl.gov/abs/hep-th/9802191}{{\tt
  hep-th/9802191}}].

\bibitem{jahn:98}
O.~Jahn and F.~Lenz, {\it Structure and dynamics of monopoles in axial gauge
  QCD},  {\em Phys.~Rev.} {\bf D58} (1998) 085006,
  [\href{http://xxx.lanl.gov/abs/hep-th/9803177}{{\tt hep-th/9803177}}].

\bibitem{ford:99a}
C.~Ford, T.~Tok, and A.~Wipf, {\it Abelian projection on the torus for general
  gauge groups},  {\em Nucl. Phys.} {\bf B548} (1999) 585,
  [\href{http://xxx.lanl.gov/abs/hep-th/9809209}{{\tt hep-th/9809209}}].

\bibitem{ford:99b}
C.~Ford, T.~Tok, and A.~Wipf, {\it SU(n) gauge theories in Polyakov gauge on
  the torus},  {\em Phys. Lett.} {\bf B456} (1999) 155,
  [\href{http://xxx.lanl.gov/abs/hep-th/9811248}{{\tt hep-th/9811248}}].

\bibitem{vandersijs:97}
A.~J. van~der Sijs, {\it Laplacian Abelian projection},  {\em Nucl.~Phys.~B
  (Proc.~Suppl.)} {\bf 53} (1997) 535,
  [\href{http://xxx.lanl.gov/abs/hep-lat/9608041}{{\tt hep-lat/9608041}}].

\bibitem{vandersijs:98}
A.~J. van~der Sijs, {\it Abelian projection without ambiguities},  {\em
  Prog.~Theor.~Phys.~Suppl.} {\bf 131} (1998) 149,
  [\href{http://xxx.lanl.gov/abs/hep-lat/9803001}{{\tt hep-lat/9803001}}].

\bibitem{vandersijs:99}
A.~J. van~der Sijs, {\it Laplacian Abelian projection: Abelian dominance and
  monopole dominance},  {\em Nucl.~Phys.~Proc.~Suppl.} {\bf 73} (1999)
  548--550, [\href{http://xxx.lanl.gov/abs/hep-lat/9809126}{{\tt
  hep-lat/9809126}}].

\bibitem{ilgenfritz:00}
E.-M. Ilgenfritz, S.~Thurner, H.~Markum, and M.~M{\"u}ller-Preussker, {\it
  Monopole characteristics in various abelian gauges},  {\em Phys. Rev.} {\bf
  D61} (2000) 054501, [\href{http://xxx.lanl.gov/abs/hep-lat/9904010}{{\tt
  hep-lat/9904010}}].

\bibitem{bali:96}
G.~S. Bali, V.~Bornyakov, M.~M{\"u}ller-Preussker, and K.~Schilling, {\it Dual
  superconductor scenario of confinement: A systematic study of Gribov copy
  effects},  {\em Phys. Rev.} {\bf D54} (1996) 2863--2875,
  [\href{http://xxx.lanl.gov/abs/hep-lat/9603012}{{\tt hep-lat/9603012}}].

\bibitem{deforcrand:01b}
P.~de~Forcrand and M.~Pepe, {\it Center vortices and monopoles without lattice
  Gribov copies},  {\em Nucl.~Phys.} {\bf B598} (2001) 557--577,
  [\href{http://xxx.lanl.gov/abs/hep-lat/0008016}{{\tt hep-lat/0008016}}].

\bibitem{reinhardt:00}
H.~Reinhardt and T.~Tok, {\it Abelian and center gauges in continuum
  Yang-Mills-theory},  \href{http://xxx.lanl.gov/abs/hep-th/0009205}{{\tt
  hep-th/0009205}}.

\bibitem{deforcrand:01a}
P.~de~Forcrand and M.~Pepe, {\it Laplacian gauge and instantons},  {\em
  Nucl.~Phys.~Proc.~Suppl.} {\bf 2001} (94) 498--501,
  [\href{http://xxx.lanl.gov/abs/hep-lat/0010093}{{\tt hep-lat/0010093}}].

\bibitem{bruckmann:01a}
F.~Bruckmann, T.~Heinzl, T.~Vekua, and A.~Wipf, {\it Magnetic monopoles vs.
  Hopf defects in the Laplacian (Abelian) gauge},  {\em Nucl.~Phys.} {\bf B593}
  (2001) 545--561, [\href{http://xxx.lanl.gov/abs/hep-th/0007119}{{\tt
  hep-th/0007119}}].

\bibitem{ryder:80}
L.~H. Ryder, {\it Dirac monopoles and the Hopf map $S^3\rightarrow S^2$},  {\em
  J.~Phys.} {\bf A13} (1980) 437--447.

\bibitem{hopf:31}
H.~Hopf, {\it {\"U}ber die abbildungen der dreidimensionalen sph{\"a}re auf die
  kugelfl{\"a}che},  {\em Math.~Ann.} {\bf 104} (1931) 637.

\bibitem{nakahara:90}
M.~Nakahara, {\em Geometry, Topology and Physics}.
\newblock Adam Hilger, 1990.

\bibitem{dubrovin:85}
B.~A. Dubrovin, A.~T. Fomenko, and S.~P. Novikov, {\em Modern Geometry -
  Methods and Applications}.
\newblock Springer, 1985.

\bibitem{deforcrand:xx}
P.~de~Forcrand. private communication.

\bibitem{brower:97b}
R.~C. Brower, K.~N. Orginos, and C.-I. Tan, {\it Magnetic monopole loop for the
  Yang-Mills instanton},  {\em Phys.~Rev.} {\bf D55} (1997) 6313,
  [\href{http://xxx.lanl.gov/abs/hep-th/9610101}{{\tt hep-th/9610101}}].

\bibitem{bruckmann:00a}
F.~Bruckmann, T.~Heinzl, T.~Tok, and A.~Wipf, {\it Instantons and Gribov copies
  in the maximally Abelian gauge},  {\em Nucl.~Phys.} {\bf B584} (2000)
  589--614, [\href{http://xxx.lanl.gov/abs/hep-th/0001175}{{\tt
  hep-th/0001175}}].

\bibitem{arafune:75}
J.~Arafune, P.~G.~O. Freund, and C.~J. Goebel, {\it Topology of Higgs fields},
  {\em J.~Math.~Phys.} {\bf 16} (1975) 433--437.

\bibitem{dirac:48}
P.~A.~M. Dirac, {\it The theory of magnetic poles},  {\em Phys.~Rev.} {\bf 74}
  (1948) 817--830.

\bibitem{taubes:84a}
C.~H. Taubes, {\it Morse theory and monopoles: Topology in long-ranged forces},
  . in: \textit{Progress in gauge field theory}, G.~'t~Hooft, ed., Plenum
  Press, New York, 1984.

\bibitem{garciaperez:00}
M.~{Garcia~Perez}, T.~G. Kovacs, and P.~van Baal, {\it Overlapping instantons},
   \href{http://xxx.lanl.gov/abs/hep-ph/0006155}{{\tt hep-ph/0006155}}.

\bibitem{hansen:xx}
D.~Hansen {\em et.~al.} work in progress.

\bibitem{bornyakov:92}
V.~G. Bornyakov, V.~K. Mitrjushkin, and M.~M{\"u}ller-Preussker, {\it
  Deconfinement transition and Abelian monopoles in SU(2) lattice gauge
  theory},  {\em Phys. Lett.} {\bf B284} (1992) 99--105.

\bibitem{ivanenko:93}
T.~L. Ivanenko, A.~V. Pochinsky, and M.~I. Polikarpov, {\it Condensate of
  Abelian monopoles and confinement in lattice gauge theories},  {\em Phys.
  Lett.} {\bf B302} (1993) 458--462.

\end{thebibliography}\endgroup

\end{document}